\documentclass{article}
\usepackage{arxiv}

\usepackage[utf8]{inputenc}
\usepackage{todonotes}
\usepackage{array}
\usepackage{mathtools}
\usepackage{subcaption}
\usepackage{enumitem}
\usepackage{cleveref}
\usepackage{subcaption}
\usepackage{url}

\usepackage{pifont}

\usepackage{blindtext}

\AtBeginDocument{%
  \providecommand\BibTeX{{\rm B\kern-.05em{\sc i\kern-.025em b}\kern-.08em
    T\kern-.1667em\lower.7ex\hbox{E}\kern-.125emX}}}

\newcommand{\newlineauthors}{%
  \end{@IEEEauthorhalign}\hfill\mbox{}\par
  \mbox{}\hfill\begin{@IEEEauthorhalign}
}

\title{A Signal Injection Attack Against Zero Involvement Pairing and Authentication for the Internet of Things}

\author{
{\rm Isaac Ahlgren}\\
Loyola University Chicago\\Computer Science Department
\and
{\rm Jack West}\\
University of Wisconsin\\Computer Science Department
\And
{\rm Kyuin Lee}\\
University of Houston\\Department of Information Science Technology
\and
{\rm George K Thiruvathukal}\\
Loyola University Chicago\\Computer Science Department
\And
{\rm Neil Klingensmith}\\
Loyola University Chicago\\Computer Science Department
}

\begin{document}

\maketitle

\begin{abstract}
Zero Involvement Pairing and Authentication (ZIPA) is a promising technique for autoprovisioning large networks of Internet-of-Things (IoT) devices. 
In this work, we present the first successful signal injection attack on a ZIPA system.
Most existing ZIPA systems assume there is a negligible amount of influence from the unsecured outside space on the secured inside space.
In reality, environmental signals do leak from adjacent unsecured spaces and influence the environment of the secured space.
Our attack takes advantage of this fact to perform a signal injection attack on the popular Schurmann \& Sigg algorithm.
The keys generated by the adversary with a signal injection attack at 95 dBA is within the standard error of the legitimate device.

\end{abstract}

\section{Introduction}

Internet of Things (IoT) devices require secure wireless communication channels to exchange data and coordinate with one another.
Without secured channels, IoT devices are susceptible to attacks such as man-in-the-middle~\cite{survdevpair, caveateptor} which jeopardize user privacy and trustworthiness of IoT devices.
To establish a secured channel, IoT devices must pair and establish a common cryptographic key.

Traditionally, IoT devices in a network individually pair with a central entity (such as a gateway or hub) that is assumed to be trusted.
This is usually done through a person intervening to type in a password.
However, human-mediated pairing is prone to many faults, particularly for IoT devices.

Passwords are a point of weakness for authentication systems in general but are particularly problematic for IoT devices.
IoT devices lack the peripherals such as a keyboard and mouse that facilitate password entry on other device classes like mobile phones, tablets, and laptops.
Furthermore, generating strong passwords, storing them securely, and periodically rotating them have all proven to be stubbornly difficult tasks for humans~\cite{lastpass, password-rotation, webpasswordhabits, nudge, soundproof}.
Thus, readily changing or updating passwords for IoT devices is tedious for users to do.
Rotating passwords on a large network of IoT devices means manually re-pairing every device by hand.
As a result, networks of IoT devices tend to be difficult to manage at scales beyond a handful of devices.

\begin{figure}
    \centering
    \includegraphics[width=\hsize]{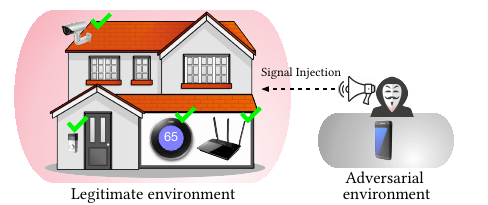}
    \caption{Legitimate devices within the same environment authenticate. External adversary authenticates by injecting a predictable signal into the legitimate environment.}
    \label{fig:aliceandbob}
\end{figure}

A central entity also presents unique problems for IoT devices.
Central hubs like WiFi access points or data concentrators can suffer temporary or permanent failure due to a malfunction during normal operations, leaving devices with no way to communicate with each other~\cite{hubmalfunc}.
Since IoT devices are often unattended (as opposed to laptops, which have dedicated users), outages can go unnoticed for long periods of time, leading to extended service interruptions.
A trusted central entity can also be compromised as a result of software vulnerabilities~\cite{hubvuln1, hubvuln2}.

As a solution to these issues, zero involvement pairing and authentication (ZIPA) aims to alleviate the need for human intervention and central entities by allowing devices to establish secure decentralized networks.
ZIPA is desirable because (1) it is always-on, (2) it is easily scalable (easy to add devices to network), and (3) it is adaptable to support mutual automatic pairing among devices from different vendors.

Devices authenticating to a ZIPA network validate their legitimacy by proving that they are located in the same physical space (i.e., office, home) at the same time. 
The devices generate authentication keys from ambient environmental contexts such as electromagnetic radiation, audio, voltage, etc, which are chosen to be observable only within the confines of a protected physical space.
Compared to traditional pairing methods, ZIPA is more secure and easier to use. 
Because the devices autonomously authenticate themselves, the user does not have to manage, remember or enter the password on individual devices.
This enhanced usability also improves the system's overall security because it allows the constituent devices to autonomously and periodically rotate keys.

However, all ZIPA schemes that we are aware of assume that there is a negligible amount of influence from the environment outside the secured space~\cite{h2h,h2b,voltkey,aerokey,ivpair,Schurmann-TMC13}.
In reality, most spaces are not impenetrable.
Environmental signals do leak from the surrounding adjacent unsecured spaces and influence the secured space's context.
Adversaries can influence the keys that ZIPA devices generate by injecting a strong known signal into the secured space from outside (depicted in Figure \ref{fig:aliceandbob}).
This type of attack is known as a \emph{signal injection attack}.

In this work, we introduce signal injection attacks on the  Schurmann \& Sigg algorithm~\cite{Schurmann-TMC13}, which is commonly used in ZIPA systems~\cite{miettinen,nguyen}.
Schurmann \& Sigg generates keys from ambient audio, which is easy to work with in a proof-of-concept system.
While we are focusing on signal injections using audio, we think that signal injection attacks can work on many different modes of signals.


Our attack is straightforward for an adversary to carry out.
Equipped with only a microphone, speaker, and laptop, the adversary sits just outside the legitimate space and broadcasts the signal toward the legitimate ZIPA devices.
The adversary then masquerades as a legitimate ZIPA device and attempts to pair.

Our attack has a high success rate for the adversary.
Keys generated by honest devices inside the legitimate space are heavily influenced by the injected signal.

Our contributions include:
\begin{itemize}

    \item We introduce a signal injection attack on the Schurmann \& Sigg algorithm that exploits the weak barriers of the room to inject a signal.
    \textbf{To our knowledge, this is the first successful signal injection attack on a ZIPA system.}

    \item We evaluate our signal injection attack on a testbed in our department's offices.
    The bit error rate of our attack's adversarial \textbf{keys are within standard error of legitimate keys}. 

    \item We suggest a mitigation for our signal injection attack.
    We leave the implementation of our mitigation to future work.
\end{itemize}

\section{Background \& Related Work}

ZIPA systems assume that devices located within the same environment (i.e., home, office) are legitimate and any devices outside of such environment is considered as an adversary, as depicted in \Cref{fig:aliceandbob}.
The physical space can be protected by conventional barriers like keycards or locks, and devices inside derive their legitimacy from being able to prove that they are nearby one another inside one of these protected areas.

\begin{figure}
    \centering
    \includegraphics[width=\hsize]{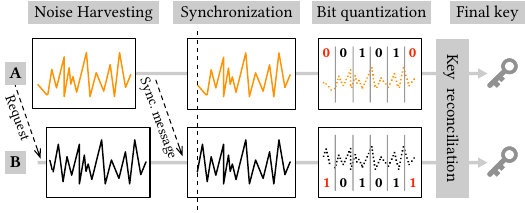}
    \caption{General pipeline of ZIPA between two authenticating
devices: $A$  and $B$ .}
    \label{fig:background}
\end{figure}

\subsection{ZIPA Pipeline Overview}

ZIPA authentication between two devices ($A$ and $B$) generally involve four stages: noise harvesting, synchronization, bit quantization, and key reconciliation as illustrated in \Cref{fig:background}.

\begin{enumerate}[leftmargin=0cm,itemindent=.5cm,labelwidth=\itemindent,labelsep=0cm,align=left]

    \item \textbf{Noise Harvesting:} Two devices independently measure a sequence of samples from an environmental signal source.
    This is usually done by a microcontroller with an on-board analog-to-digital converter or other sensors.
    
    \item \textbf{Synchronization:} 
    Two devices time-align their signal measurements with one another.
    One device sends a short snippet of the measurement results (synchronization message) to the second device over a public unsecured channel.
    The second device shifts the synchronization message over its own measured signal to obtain an identical starting point by searching for maximized Pearson correlation coefficient.
    
    \item \textbf{Bit quantization:} \label{step:quant}
    Both devices independently convert the synchronized signal into a bit sequences to be later used as a key.
    
    \item \textbf{Key Reconciliation:} 
    The two devices exchange messages with one another over a public channel to resolve bit differences in their quantized bit sequences.
\end{enumerate}

\subsubsection{Schurmann \& Sigg Bit Quantization}
\label{sec:sigg}
\begin{figure*}[t]
    \centering
    \includegraphics[width=\hsize]{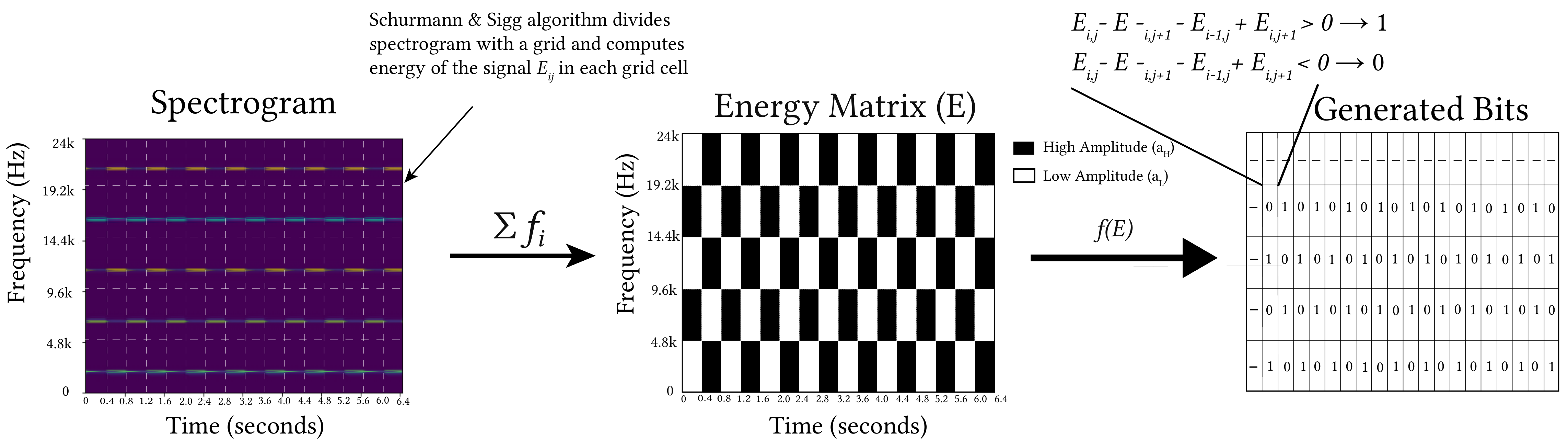}
    \caption{Overview of our attack on the Sigg algorithm. The spectrogram (left) is divided in a grid. The energy of each box in the grid is computed to build an energy matrix $E$.  Bits are computed as a function of nearby cells on $E$.}
    \label{fig:inj_sig_attack}
\end{figure*}
Our attack targets the popular Schurmann and Sigg bit quantization algorithm, which generates a sequence of bits from a time series of environmental sensor readings (step \ref{step:quant} in the sequence above). 
The steps to generate a bit sequence are illustrated in Figure \ref{fig:inj_sig_attack}.

\begin{enumerate}
\item First, the algorithm builds a spectrogram from the time series samples of the environmental signal.

\item The spectrogram is divided into a grid by frequency and time.
The algorithm computes the energy of each cell in the spectrogram's grid, assembling this results in a matrix (denoted by $E$).

\item The bit quantization algorithm used to generate the keys for the Schurmann and Sigg algorithm utilizes the difference in energy between frequency bands in the spectrogram.
The bit quantization is described as:
\begin{equation}
\label{eqn:siggquant}
f(E) =
  \begin{cases}
      1 & (E_{ij} - E_{i j+1} - 
             (E_{i-1j} - E_{i-1j+1})) > 0 \\
      0 & otherwise \\
  \end{cases}
\end{equation}
\end{enumerate}

where $E$ is the energy matrix of the spectrogram, $i$ represents the index for the frame, and $j$ represents the index for the frequency band.

\subsection{Related Work}

\label{sec:realted_work}
Our work focuses on the on stronger adversarial attacks on ZIPA systems.
Most ZIPA systems assume a weaker adversary for their threat model and thus perform weaker attacks on their systems.
Generally, threat models from prior work can be categorized in two ways: a passive adversary and a passive-active adversary.

For passive adversary threat model, the adversary watches and observes the context coming from the secured space from an unsecured space.
Typically, its assumed that an adversary only has access to the context observed from the unsecured space and does not try to actively try to manipulate the authentication process~\cite{Miettinen-CCS14,Saxena-S&P06,Mathur-Mobisys11}.

For the passive-active threat model, the adversary attempts to actively manipulate the authentication process through attacks such as replay attacks and machine learning attacks~\cite{aerokey,voltkey}.
A replay attack is where an adversary records the environmental context at an earlier time and replays it at a later time to masqurade as a legitimate device.
The SoundDanger attack~\cite{sounddanger}, the attacker induces a sound in the legitimate space, for example by sending a text message or making a phone call that causes a device to produce a known sound.
A machine learning attack is where a machine learning model is trained on environmental context collected at an earlier time to predict the environmental context later.
These threat models assume a stronger adversary, however this adversary is still unmotivated and weaker compared to our attack.

AeroKey~\cite{aerokey} is the only ZIPA system we know of that attempts a signal injection attack.
However, their signal injection attack was unsuccessful.
This could be due to a low injection signal intensity that cannot overwhelm the barriers of the secured space.
Also, the type of signal injected in AeroKey may not have been a good fit for the system's bit quantization algorithm.

\subsection{Threat Model}
We assume that legitimate devices are all located inside some physically-secured space such as an office, house, or apartment.
We call this space the legitimate environment.
Individual devices may have limited computation power---IoT devices, mobile phones, or bluetooth accessories all have microcontrollers or low-power CPUs.

Adversaries are restricted from the physically-secured space, but they may have access to adjacent rooms like hallways or neighboring apartments.
These neighboring spaces are the adversarial environment.
They can intercept all public (unsecured) wireless transmissions made by the legitimate devices inside the legitimate environment.
Although they do not have access to the same ambient environmental signals inside the space, they can measure the environmental context nearby, which is only slightly different from the contexts of legitimate environment as illustrated in \Cref{fig:aliceandbob}.
Adversaries can also masquerade as legitimate devices by initiating authentication. 
Additionally, although the legitimate device's computation power may be limited, we assume the adversary has much more computation power at its disposal (eg. GPUs).

The adversary's goal is to gain access to the network of legitimate devices either by (a) passively listening to wireless communications and environmental signal to learn their shared key or by (b) actively masquerading and participating in the protocol as a legitimate device.
To learn the shared key, the adversary passively snoops the public wireless channel for messages exchanged by legitimate devices.
The information it learns by listening to the channel may be used to form an estimate of the key.
In the masquerading strategy, the adversary impersonates a legitimate device, acquiring measurements of ambient environmental signals and following the protocol to establish a key.
Using a combination of imperfect measurements of the environmental signal and messages exchanged with legitimate devices, the adversary negotiates a key and joins the network

\section{Signal Injection Attacks}
\label{sec:problem_validation}

In an ideal setting, legitimate devices on a ZIPA network are isolated from external adversaries by a barrier that does not allow environmental signals to pass.
A locked soundproof room, for example, could perfectly isolate audio from the outside world.
Malicious devices outside the isolated room cannot hear the environmental context inside, and they cannot influence the environmental context inside.
Without access to the environmental context, outside devices cannot generate a valid key.
The threat model of existing ZIPA systems presumes near-perfect isolation~\cite{aerokey,Mathur-Mobisys11,Miettinen-CCS14,Schurmann-TMC13,voltkey}.

In reality, the boundaries that separate physical spaces are much more pourous to environmental noise.
Many of us have experienced loud noises coming from outside a building thar are audible inside a room.
Other forms of environmental context like electromagnetic radiation, ambient light, and others can analogously be influenced from the outside.
Our evaluation confirms that environmental context is detectable outside the enclosed legitimate space.

Our signal injection attack takes advantage of the imperfect barrier between the enclosed legitimate environment and the outside world.
We ask the question: \emph{is it possible for an external adversary to broadcast a signal from outside the legitimate environment that causes legitimate ZIPA devices to produce a known key?}

If so, an adversary can control the key generation process and gain unauthorized access to a ZIPA network.
In this section, we describe  a signal injection attack for the popular Schurmann and Sigg bit quantization algorithm using audio as a source of environmental context.
We demonstrate both a simulated and live signal injection attack on the Schurmann \& Sigg algorithm.

\subsection{Description of Our Signal Injection Attack}
To experiment with signal injection attacks, we set up a ZIPA testbed depicted in \Cref{fig:signal_injection}.
Inside the legitimate space, we placed a legitimate ZIPA device next to a television playing a video for environmental context.
Outside the legitimate space, we placed a second ZIPA device along with a speaker to generate an artificial signal that overwhelms the legitimate context from the TV.

Our artificial injection signal is engineered to produce a predictable key when used as input to the Schurmann \& Sigg bit quantization algorithm.
The injection signal sufficiently overwhelms the context in the legitimate space, and an adversary can break the key with relative ease.
The challenge is to decide what signal we should inject to induce a desired key.

\subsection{Signal Injection Attack on the Schurmann and Sigg Algorithm}

\begin{figure}
    \centering
    \includegraphics[width=\hsize]{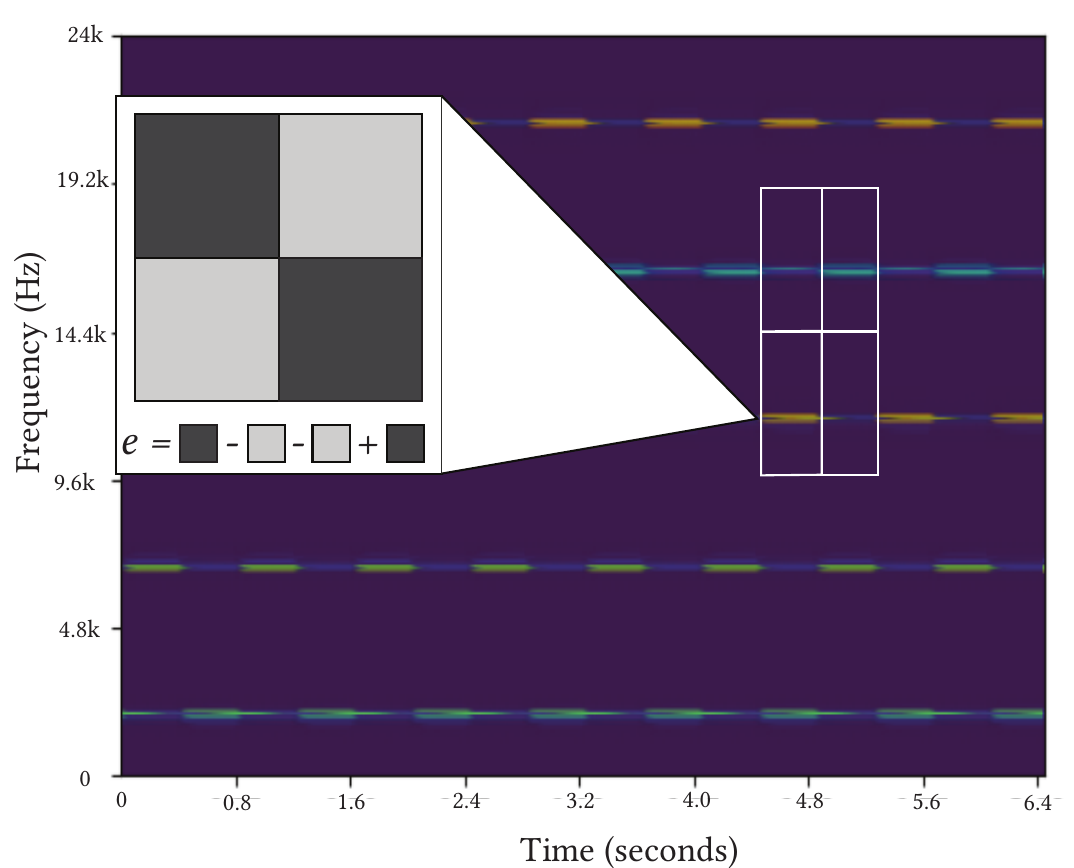}
    \caption{Spectrogram of our injection signal with a blown up view of the Schurmann \& Sigg grid that shows how elements of the energy matrix are computed from the spectrogram.}
    \label{fig:siggsquare}
\end{figure}

The Schurmann and Sigg bit quantization algorithm (\S \ref{sec:sigg}) produces bits based on the energy differences of adjoining frequency bands in a spectrogram, as depicted in \Cref{fig:siggsquare}.
The algorithm divides the spectrogram with a grid of divisions in frequency and time.
It then compares the energy of adjacent rectangles in the grid, assigning bits according to \Cref{eqn:siggquant}.
We want to design an injection signal that induces the quantization algorithm to generate a predictable pattern of 1's and 0's.

\begin{itemize}[leftmargin=0cm,itemindent=.5cm,labelwidth=\itemindent,labelsep=0cm,align=left]
\item \textbf{Inducing 1's:} We can induce a \textbf{positive} value for $e$ by setting the dark squares to have \textbf{large} positive energy values $a_H$ while the light squares have \textbf{small} or zero energy $a_L$.
The Schurmann and Sigg algorithm computes:
\begin{equation}
\label{eqn:e}
e = a_H - a_L - a_L + a_H
\end{equation}
Choosing $a_L = 0$ forces $e$ to be positive:
$$
e = a_H - 0 - 0 + a_H = 2a_H > 0
$$
And bit quantization produces a $1$ for positive values of $e$.
\item  \textbf{Inducing 0's:} We can induce a \textbf{negative} value for $e$ by setting the dark squares to have \textbf{small} or zero energy values $a_L$ while the light squares have \textbf{large} positive energy values $a_H$:
The Schurmann and Sigg algorithm computes:
$$
e = a_L - a_H - a_H + a_L =  0 - a_H - a_H + 0 = -2a_H < 0
$$
The Schurmann and Sigg algorithm produces a $0$ for negative values of $e$.
\end{itemize}

We build our injection signal by creating a spectrogram that has tiled checkerboard of high-energy (dark) and low-energy (light) rectangles.
High-energy rectangles have a high-amplitude sine wave.
Low-energy rectangles have no signal.

\section{Evaluation}
\label{sec:eval}

\begin{figure}
    \centering
    \includegraphics[width=\hsize]{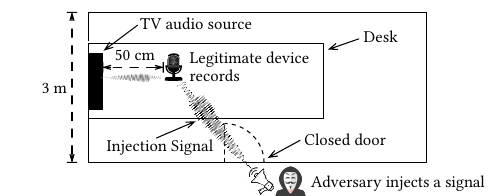}
    \caption{Illustration of the testbed we used to carry out our injection. Adversary sits outside an office with the door closed and broadcasts the injection signal into the legitimate space. Legitimate device inside the office pairs with adversary.}
    \label{fig:signal_injection}
\end{figure}

We evaluated our injection attack in a testbed of ZIPA devices located in our department's offices.
The environment for the experiment is depicted in \Cref{fig:signal_injection}.

In our testbed, there are two legitimate devices inside of a secured space and one adversarial device in a publicly-accessible hallway immediately outside.
All devices are using a HyperX SoloCast USB microphone~\cite{solocast} to record audio at a sampling rate of 48kHz.
The adversarial device uses a Samsung HW-J355 soundbar~\cite{speaker} as a speaker to inject the injection signal into the secured space.
The injection signal was evaluated using two methods: a simulation and a real testbed.
The simulation evaluated the injection signal in perfect signal propagation conditions.
The testbed  evaluated the injection signal in real life signal propagation conditions.

For both methods, we played a YouTube video\footnote{\url{https://youtu.be/HeQX2HjkcNo?si=bqW6vsmvnv2fIsCh}} on a TV in our testbed to simulate conversation in the room.
We synchronously recorded audio inside the room and outside the room for approximately 30 minutes.
For the simulation, we superimposed our injection signal onto the environmental context by adding the two signals together in software.
For the testbed, a signal was injected by the adversary using a speaker in which the signal's intensity varied over consecutive trials of the experiment.
Since the simulation does not experience distortion from a real environment, it is a good approximation of the best performance our attack can achieve under ideal conditions.

\subsection{Signal Injection Influence}
\label{sec:influence}

\begin{figure}
    \centering
    \includegraphics[width=\hsize]{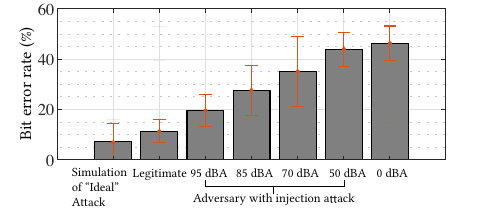}
    \caption{Comparison of bit error rates of keys generated by a pair of legitimate devices and adversary (lower is better). Adversary using injection attack at high volume achieves nearly the same bit agreement rate as legitimate device.}
    \label{fig:barplot}
\end{figure}

The effectiveness of our signal injection attack is illustrated in \Cref{fig:barplot}.

Without issues of signal propagation in the simulation, the attack generated keys within an average of $\sim$7\% bit error rate of the keys generated within the secured space.
This is only slightly better than the bit error rate of the keys generated by devices within the secured space at an average bit error rate of $\sim$11\%.

On the testbed, a signal injection attack at 95 dBA had an average bit error rate of $\sim$20\%.
The signal injection attack at 85 dBA had an average bit error rate of $\sim$27\%.
The signal injection attack at 70 dBA had an average bit error rate of $\sim$35\%.
The signal injection attack at 50 dBA had a similar effect to no injection signal at all as it had an average bit error rate of $\sim$44\% while no injection attack had an average bit error rate of $\sim$46\%.
The intensity of the injection signal scales directly proportional to the decreased bit error rate.
Hence, we can conclude that a higher intensity signal can decrease the bit error rate even more so.

\subsection{Injection Signal Phase}

\begin{figure}
    \centering
    \includegraphics[width=\hsize]{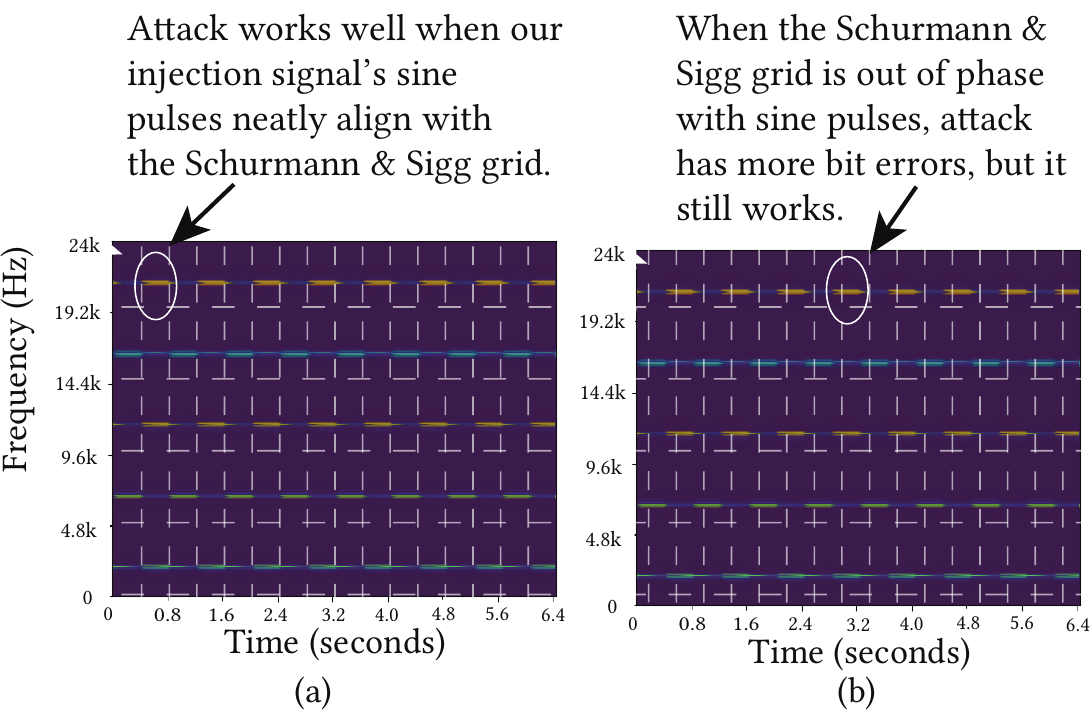}
    \caption{Phase alignment of our injection signal with the Schurmann \& Sigg grid.}
    \label{fig:phase}
\end{figure}

In general, ZIPA bit quantization algorithms tend to perform well only when they generate keys from time-aligned buffers.
A pair of legitimate ZIPA devices must synchronize their sample buffers to ensure the bit quantization algorithms produce the same bit sequence from their measurements of environmental context.

Our injection attack superimposes an artificial signal over the legitimate space's environmental context.
The attack works best when the sine waves of the injected signal's spectrogram are perfectly aligned with the Schurmann \& Sigg grid, as depicted in \Cref{fig:phase}(a).
If the injection signal is out of phase with the legitimate device's grid (as in \Cref{fig:phase} (b)), some energy from dark squares in \Cref{fig:siggsquare} will bleed into the light squares.
This can cause the calculation of $e$ in \Cref{eqn:e} to be less than $2a_H$, which makes the injected signal less reliable at producing 1's and 0's in a predictable sequence.

\begin{figure}
    \centering
    \includegraphics[width=\hsize]{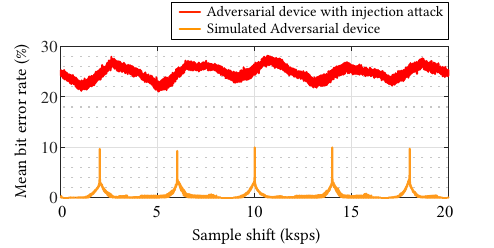}
    \caption{Plot of bit error rate as a function of shift amount for our injection attack. Shifting the adversary's injection signal relative to the legitimate device's increases bit errors, but the adversary can often still pair.}
    \label{fig:injection_attack2}
\end{figure}

We tested the effect that shifting the injection signal has over the overall bit agreement rate of the generated key, shown in \Cref{fig:injection_attack2}.
In the figure, a shifted injection signal does indeed produce higher bit error rates.
While there is no way for us to reliably time-align the injection signal with the algorithm's grid, our data shows us that the worst case bit error rate is still below 30\%---low enough for pairing to succeed.
The injection attack can still succeed even if the injected signal is out of phase with the grid.

\subsection{Generated Key Entropy}
While the injection signal can produce a mostly predictable set of bits for the legitimate devices key, it raises the question if the a legitimate device can detect the attack taking place through observing the apparent predictability of its generated keys.
Using the 95 dBA injection signal from our testbed experiment, we generated 12,300 bits to calculate the entropy of the generated keys.
The entropy was calculated using symbols of size 8 bits.
The calculated entropy from the bits came to \textbf{$\sim$7 bits of entropy}.
Hence, we conclude that the attack is difficult to detect from key entropy, as the entropy is not suspiciously low.



\section{Discussion \& Future Work}


Our attack uses a specially-tailored injection signal to target the Schurmann \& Sigg bit quantization algorithm.
But we suspect injection attacks against other ZIPA systems would be successful with other signals that are not tailored to a specific bit quantization algorithm if the amplitude was high enough to overwhelm the context in the legitimate space.
For ZIPA systems that use audio as environmental context, a signal injection attack would be obvious to people in the legitimate environment because they will hear the sound being broadcast.
In other forms of environmental context like electromagnetic radiation, an injection attack may not be easily detectable.
We need some technique to prevent these attacks if ZIPA systems are to be practical.

Devices on a ZIPA network establish their legitimacy from signals measured in a physically-restricted space.
The key observation of  this work is that it is possible for an external attacker to influence the legitimate environmental context.
So ZIPA systems that rely solely on a shared environmental signal are prone to injection attacks.

\subsection{Detecting Injection Attacks}

In general, detecting injection attacks against ZIPA networks is a difficult problem and depends on the attacker's strategy.

\begin{itemize}
\item \textbf{Anomoly Detection:} many techniques exist for detecting signal anomalies~\cite{anomalydetection}.
It is possible that some of these techniques could be used to detect signal injection attacks.
\item \textbf{Online NIST Test for Randomness:} we expect injection attacks to produce bit sequences with relatively low randomness.
We did a simple test for entropy, but we did not pass bit sequences through the NIST test~\cite{NIST} because our experiments did not generate enough bits to produce a conclusive result.
By continuously evaluating the randomness of bit sequences in real time as they are produced, it may be possible to identify signal injection attacks.
This may be challenging because the NIST test codebase is large and not designed to run on low-resource embedded computers that are typically used on ZIPA devices.
\end{itemize}

\subsection{Mitigation using Impulse Responses}

This work demonstrates that it is easy to spoof real-time measurements of ambient environmental signals.
We think a mitigation technique for injection attacks would combine properties of the physical space by using an impulse response as context for key generation.
As impulse responses don't rely on the ambient environment signals, real-time measurements of these properties cannot be spoofed by a signal injection attack. 

\begin{figure}
    \centering
    \includegraphics[width=\hsize]{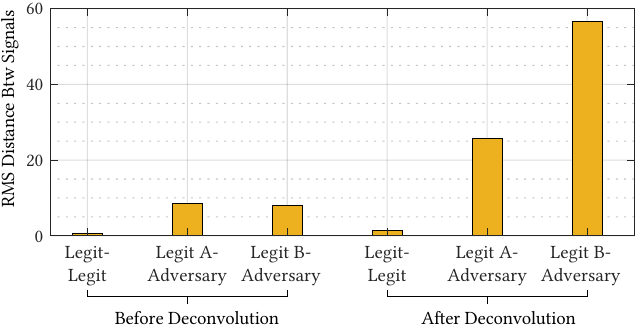}
    \caption{Deconvolving environmental signal with room impulse response amplifies the difference between legitimate devices and adversaries. Bigger RMS distance between legitimate and adversary is better.}
    \label{fig:deconvolve}
\end{figure}

We demonstrate this concept with a preliminary experiment.
First, we estimated the impulse response of our testbed by playing a sine sweep~\cite{farina2000simultaneous} on one legitimate device and recording on a nearby legitimate device in the same room.
We used the same technique to measure the impulse response between a legitimate device and a distant adversary located in the publicly accessible hallway outside our testbed.

We then synchronously recorded ambient audio (played on the TV in the testbed) on all three devices.
As we expected, the recordings of the two legitimate devices were very similar, and the recording of the adversary was slightly different in RMS distance.
\Cref{fig:deconvolve} shows the RMS distance between environmental signals.
Deconvolving the room impulse response from the ambient audio recordings substantially amplified differences between the legitimate and adversary devices.

The intuition behind why this deconvolution works is that impulse response estimates generally contain a lot of noise.
In our testbed, the building's HVAC system, computer fans, and other extraneous sounds all add noise to the impulse response, even for a pair of nearby devices.
The adversary, located outside the legitimate space, records an impulse response with a very low signal to noise ratio (SNR).
Noise in the adversary's impulse response propagates through the deconvolution to produce a context that is very different from the legitimate device.

\section{Conclusion}

In this work, we presented the first successful signal injection attack on a ZIPA system to our knowledge.
We studied the Schurmann \& Sigg algorithm to engineer a signal that made it produce predictable keys when injected into the secured environment.
We evaluated our injection signal over simulation and through a live testbed.
On our testbed, the attack produces keys that are statistically similar to keys produced by the legitimate devices.
To help against this attack we proposed a couple of potential mitigation techniques.
We illustrated some preliminary results for a particular mitigation technique based on room impulse responses.
The result showed the feasbility of using room impulse responses as a contextual feature for ZIPA systems.

Our attack studies ZIPA systems that produce keys from environmental audio, and it leaves open the question about whether our signal injection attack would be effective against other modes of environmental signal.
Future work could study the question of whether injection attacks would be successful against ZIPA systems that use other modes of environmental noise to establish context.

\bibliographystyle{plain}
\bibliography{refs}

\end{document}